\documentclass[sigconf, 9pt]{acmart}
\usepackage{booktabs}
\usepackage{float}
\usepackage{todonotes}
\usepackage[algoruled,resetcount,linesnumbered]{algorithm2e}
\usepackage [autostyle, english = american]{csquotes}
\MakeOuterQuote{"}

\AtBeginDocument{%
  \providecommand\BibTeX{{%
    \normalfont B\kern-0.5em{\scshape i\kern-0.25em b}\kern-0.8em\TeX}}}

\setcopyright{acmcopyright}
\copyrightyear{2018}
\acmYear{2018}
\acmDOI{10.1145/1122445.1122456}

\acmConference[Woodstock '18]{Woodstock '18: ACM Symposium on Neural
  Gaze Detection}{June 03--05, 2018}{Woodstock, NY}
\acmBooktitle{Woodstock '18: ACM Symposium on Neural Gaze Detection,
  June 03--05, 2018, Woodstock, NY}
\acmPrice{15.00}
\acmISBN{978-1-4503-9999-9/18/06}

\SetKwInput{KwInput}{Input}
\SetKwInput{KwOutput}{Output}
\SetKwProg{Fn}{Function}{}{end}

\begin{document}

\title{Exploring Graph Based Approaches for Author Name Disambiguation}

\author{Chetanya Rastogi}
\authornote{All authors contributed equally to this research.}
\affiliation{\institution{Stanford University}}
\email{chetanya@stanford.edu}

\author{Prabhat Agarwal}
\authornotemark[1]
\affiliation{\institution{Stanford University}}
\email{prabhat8@stanford.edu}

\author{Shreya Singh}
\authornotemark[1]
\affiliation{\institution{Stanford University}}
\email{ssingh16@stanford.edu}


\begin{abstract}
  In many applications, such as scientific literature management, researcher search, social network analysis and etc, Name Disambiguation (aiming at disambiguating WhoIsWho) has been a challenging problem. In addition, the growth of scientific literature makes the problem more difficult and urgent. Although name disambiguation has been extensively studied in academia and industry, the problem has not been solved well due to the clutter of data and the complexity of the same name scenario. In this work, we aim to explore models that can perform the task of name disambiguation using the network structure that is intrinsic to the problem and present an analysis of the models.
\end{abstract}



\begin{teaserfigure}
    \centering
  \includegraphics[scale=0.4]{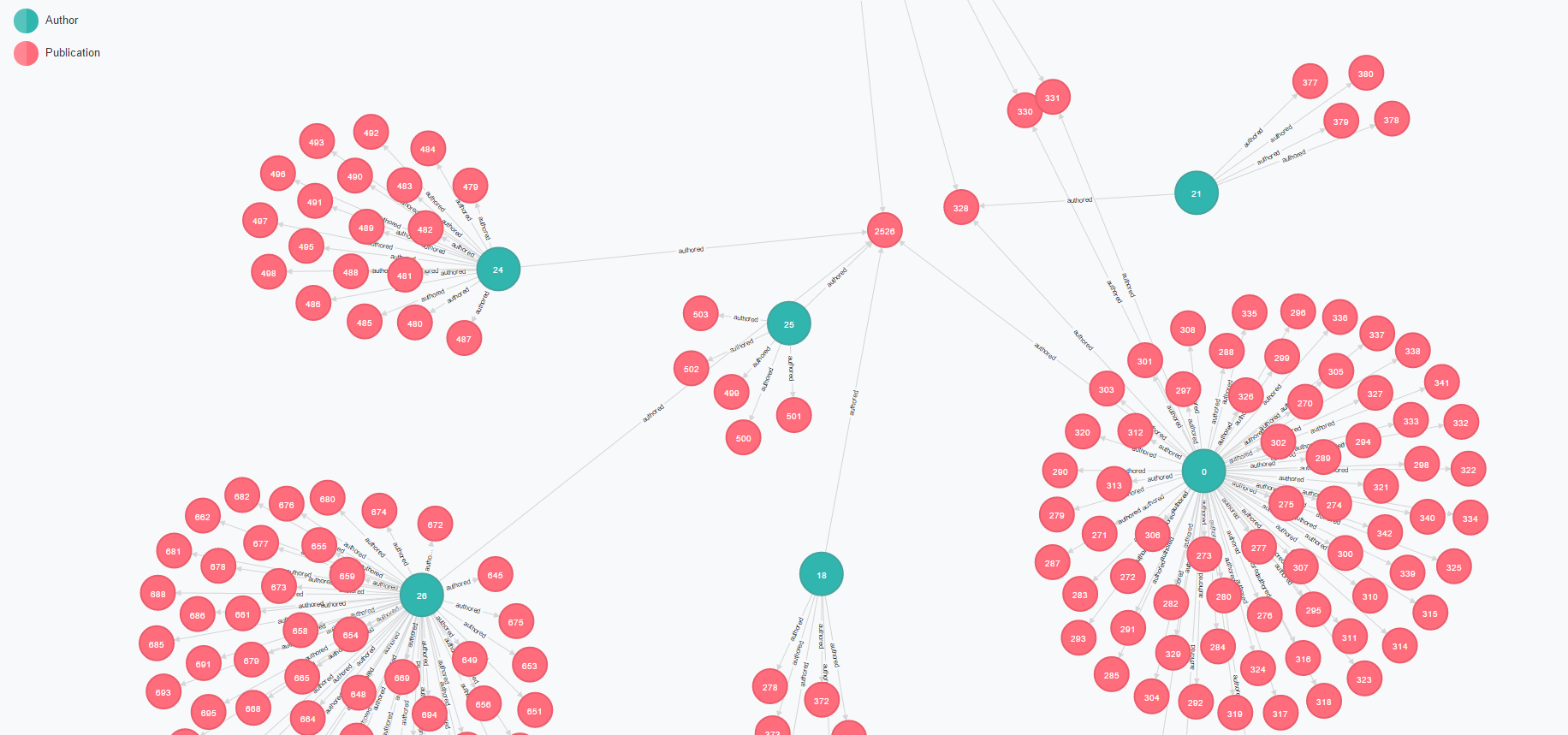}
  \caption{The ever expanding author publication network}
  \label{fig:teaser}
\end{teaserfigure}

\setcopyright{none}
\settopmatter{printacmref=false} 
\renewcommand\footnotetextcopyrightpermission[1]{} 
\pagestyle{plain} 
\maketitle

\section{Introduction}
Online academic search systems (such as Microsoft Academic Graph, Google Scholar, Dblp, and AMiner) have a large amount of research papers, and have become important and popular academic communication and paper search platforms. However, due to the limitations of the paper assignment algorithm, there are many papers assigned to error authors. In addition, these academic platforms are collecting a large number of new papers every day (AMiner has about 130,000,000 author profiles and more than 200,000,000 papers) \cite{Zhang:2018:NDA:3219819.3219859}. Therefore, how to accurately and quickly assign papers to existing author profiles, and maintain the consistency of author profiles is an urgent problem to be solved for current online academic systems and platforms. \\ 

In our project, we aim to implement author name disambiguation techniques to disambiguate profiles of authors with similar names and affiliations. We study the problem from a network perspective where researchers communicate with one another by means of their publication. The network is modeled as a bipartite graph containing two types of nodes, viz. author nodes and paper nodes. Each edge in the graph represents an author's contribution to a paper. We believe that  this inherent structure will be able to encapsulate much more implicit and intrinsic features that are otherwise impossible to capture using bibliometric data.

\section{Related Work} \label{lit_survey}
The problem of Author Name Disambiguation has been of interests to researchers for a quite long time. \cite{Han:2004:TSL:996350.996419} formulates it in the paradigm supervised learning and makes use of the various features associated with a publication, including title, co-authors, conference, etc to make the correctly associate a publication with a specific author by learning the linkage function between the publication and the author. The authors makes use of two different datasets, one from DBLP and the other collected from the web, and test two different classification algorithms on both the datasets. However, the authors do not take into account the implicit network structure that lies in the dataset. \cite{treeratpituk2009disambiguating} furthers the task and provides an extensive study on choosing a minimal subset of features by means of a random forest classifier that can identify the correct author entity linked with a particular publication. The authors also introduce a new dataset called Medline which is particular to the researchers of biomedical science. Once again the authors of these work not only ignore the underlying graph structure but also restrict the work to a particular domain which constraints the problem to a very narrow dataset. 
\\

Another important shortcoming of both the above works is that the authors know beforehand of how many clusters they need to identify for a particular name. This challenge is tackled by \cite{tang2011unified, 10.1145/3308560.3316599} as they investigate a dynamic approach for estimating the number of people associated with a particular name. They propose a novel approach of framing the problem as a Markov Random Field and try to make use of the underlying graph structure by defining similarity both in terms of the content of the publication as well as the relationship between them in terms of co-authors. Similarly, \cite{Zhang:2018:NDA:3219819.3219859} also talks about learning the cluster size dynamically and quantifying the similarity in terms of the graph structure. 
\\

In \cite{ma2019author}, Ma et al. have proposed a novel AND (Author Name Disambiguation) approach which tries to disambiguate authorship of research papers. As the population is growing, some people will inevitably share some personal features at different levels (like names and affiliations). This poses a huge challenge for many applications like information retrieval and academic network analysis. The dataset used by the authors in this work is the AMiner dataset which is a heterogeneous academic network consisting of multiple entities (i.e. author, paper, venue, topic) as well as relationships (i.e. writing, publishing, collaborating and affiliations). To solve the problem of name disambiguation, the authors propose a meta-path channel based heterogeneous network representation learning method (Mech-RL) wherein node embeddings are learned from the whole heterogeneous graph instead of breaking it down to simpler subgraphs.\\

The node (paper) embeddings \cite{singh2018footwear} are learned at two levels: they are initialized by the textual features (Doc2vec embeddings) and further optimized by relational features (from the metapaths in which they appear). Once, each entity (here paper) is represented through its low dimensional embeddings, the task is reduced to a clustering task where each cluster will contain papers belonging to a unique person. Another thing to be noted is that in this approach, the authors solve the name disambiguation problem without considering the private information of the researchers. The experimental results based on the AMiner dataset show that Mech-RL obtains better results compared to other state-of-the-art author disambiguation methods \cite{zhang2014name, schulz2014exploiting, singh2019one}.

\subsection{Bibliometric approaches to tracking international scientific migration}

Scientific migration/mobility is a well-studied topic in sociology. With the availability of large scale bibliometric data available online (Scopus, MAG, DPLB, etc.), many studies have been done to quantify scientific mobility on a large dataset. Since the academic network data is noisy and has missing data, people have used different methods to address the concerns of name disambiguation, geo-tagging, etc. Hadiji et. al. \cite{hadiji2015computer, kumari2017parallelization} uses compressed label propagation to infer missing geo-tags of
author-paper-pairs retrieved from online bibliographies like ACM, DBLP, etc. Robinson et. al. \cite{robinson2019many} used the name-disambiguation method from \cite{caron2014large} to augment the data. Moed et. al. \cite{moed2014bibliometric} used Scopus data to circumvent these noises. \\

All these studies then use statistical measures to model different aspects of migration for each author considered separately over the period considered in the study. Hadiji et. al. estimates the distribution for move propensity of an author, whereas Moed et. al. analyzes the relative migration index for 17 country pairs. Robinson et. al. introduces a new taxonomy to account for different mobility patterns rather than just migration and classifies each author to one of the class based on their affiliation history and presents an analysis of the different mobility classes for different countries.

\subsection{Characterizing evolution of graph over time}

In \cite{de2016quantifying}, Domenico et al. try to understand the dynamics of an academic network to determine the flow of authors' research interests - which they refer to as "the knowledge diaspora". They use the Microsoft Academic Graph \cite{sinha2015overview, 10.1007/978-981-10-8639-7_25} and the SCImago \cite{SCImago} classification to categorize each paper under different areas of knowledge and study the temporal snapshots to identify a growing/falling interest in a particular area of study. By studying this question from a network perspective and modeling it as a multi-layer network, the authors formulate a quantitative metric to indicate the "attractiveness" of a topic through time and are able to relate the metric with the corresponding historical or political events during that time. Furthermore, the authors also provide a metric to quantify whether a particular area of study is serving as a "source" -  supplying other areas with researchers - or a "sink" - attracting researchers  towards higher trans-disciplinary and multidisciplinary research. \\

Dynamic network analysis \cite{jindal2023classification} is a sub-field of network science aiming at representing and studying the behavior of systems constituted of interacting or related objects evolving through time. While there is substantial work in macroscopic (graph level) and mesoscopic (community level) analysis of such networks, microscopic analytic methods are less studied \cite{aggarwal2014evolutionary}. In \cite{orman2017exploring}, Orman et. al. introduces the concept of neighborhood events as a measure to characterize a node behavior across time steps. They also present a parallelizable algorithm to detect such events efficiently and show that this event sequence characterization can be used to analyze global trends in the network as well as individual node characterization:
\begin{enumerate}
    \item Node characterization: They cluster nodes based on the count of events in a time slice $t_i$ and are able to identify clusters of stable nodes and active nodes. Moreover, they observe that these clusters have different most frequent event sequences.
    \item Global trends: They used frequent pattern mining to identify certain trends among the nodes at the level of the network and found that the Enron trends reflect the routine of sporadically sending/receiving emails, whereas those of LastFM and DBLP describe a similar life cycle for ego-components: creation, growth, and decline.
\end{enumerate}
\section{Data}
\subsection{Author Name Disambiguation}
We utilise a dataset hosted as a part of a competition called OAG-WhoIsWho Track 1 \cite{biendata}. The organizers provide three different datasets for training, validation and testing of models but provide the ground truth labels for only the train set. Therefore, to test our models and provide quantitative metrics of our methods we utilise only the training set which we now refer to as the "entire" set. \\

\textbf{Task Description:} Given a bunch of papers with authors of one same name, the task is to return different clusters of papers. Each cluster has one author, and different clusters have different authors, although they have the same name.\\

\textbf{Data Description:} The dataset consists of two sets of information: list of publications for same author name and metadata of the publications. The format and fields of the publication metadata is described in Table \ref{table:data-description}. Additionally the train data contains publications of same author name, clustered by author profile which is the required output of the task. Initial data exploration on the metadata showed that the data is very noisy and has many typos and wrong entries, which makes it non-consumable in the raw form. Therefore, we pre-process and augment the data which is described in the next section. \\

We run our experiments under supervised as well as unsupervised learning paradigm. To allow for fast and feasible experimentation, we sample 20 names at random from the entire set on which we train and evaluate our methods. For unsupervised methods, we use the complete sampled dataset for training as well as evaluation while for supervised learning methods the sampled dataset is split it into train, validation and test with 15, 2, and 3 names respectively. To verify that the randomly sampled set is a valid placeholder for the entire set, we compare different attributes of the graphs generated by both and see similar distributions. The data summary comparing the statistics of the sampled set with the entire set has been shown in Table \ref{tab:sampled-comparison}. The data summary for the train, validation and test sets  has been shown in Table \ref{tab:data-stats}
\begin{table*}[t]
\begin{tabular}{@{}llll@{}}
\toprule
\textbf{Field} & \textbf{Type}   & \textbf{Meaning}   & \textbf{Example}                                                          \\ \midrule
id                                 & string                            & PaperID                              & 53e9ab9eb7602d970354a97e                                                                      \\
title                              & string                            & Paper Title                          & Data mining: concepts and techniques                                                          \\
authors.name                       & string                            & Authors                              & Jiawei Han         \\                                                             
author.org                         & string                            & Organization                         & department of computer science university of illinois at urbana champaign                     \\
venue                              & string                            & Conference/Journal                   & Inteligencia Artificial, Revista Iberoamericana de Inteligencia Artificial                    \\
year                               & int                               & Publication                          & 2000                                                                                          \\
keywords                           & list of strings                   & Key words                            & {[}"data mining", "structured data", "world wide web", "social network", "relational data"{]} \\
abstract                           & string                            & Abstract                              & Our ability to generate...                                                                    \\ \bottomrule
\end{tabular}
\caption{Description of the fields in the paper data}
\label{table:data-description}
\end{table*}

\begin{table}
\begin{tabular}{@{}lrrr@{}}
\toprule
\textbf{Parameter}    & \multicolumn{1}{p{0.2\linewidth}}{\textbf{Entire Set}} & \multicolumn{1}{p{0.2\linewidth}}{\textbf{Sampled Set}} \\ \midrule
\textbf{Distinct author names}      & 221   & 20    \\
\textbf{Distinct author profiles} & 22839 & 1945   \\
\textbf{\# of publications} & 203184 & 16788 \\ 
\textbf{\# of connected components}& 22105 & 1927 \\ 
\textbf{Largest Connected Component}& 16339 & 1048 \\
\bottomrule
\end{tabular}
\caption{Statistics for Sampled set and the Entire set}
\label{tab:sampled-comparison}
\end{table}

\subsection{Data pre-processing and summary}\label{sec:data}
\subsubsection{Size of the dataset}
\begin{figure}
    \centering
    \includegraphics[scale=0.6]{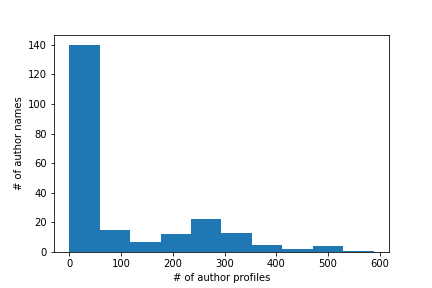}
    \caption{Author name frequency distribution by profile count}
    \label{fig:auth_profile}
\end{figure}

\begin{figure}
    \centering
    \includegraphics[scale=0.6]{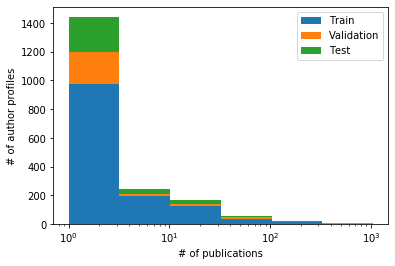}
    \caption{Cluster size distribution(number of publications per author profile)}
    \label{fig:data_stats}
\end{figure}

The number of publications, distinct author names and author profiles is shown in Table \ref{tab:data-stats} and Figure \ref{fig:data_stats} shows the distribution of number of publications across all authors profiles across the sampled dataset. There are on average 103.34 distinct author profiles for each author name in the entire dataset. The distribution of author profile count for author name is shown in Fig. \ref{fig:auth_profile}. 

\begin{table}
\begin{tabular}{@{}lrrrr@{}}
\toprule
\textbf{Dataset}    & \multicolumn{1}{p{0.16\linewidth}}{\textbf{\# of publications}} & \multicolumn{1}{p{0.12\linewidth}}{\textbf{\# of author names}} & \multicolumn{1}{p{0.16\linewidth}}{\textbf{\# of author profiles}} & \multicolumn{1}{p{0.20\linewidth}}{\textbf{avg. publications per author profile}} \\ \midrule
\textbf{Train}      & 10966  & 15 & 1347 &   8.14     \\
\textbf{Validation} & 1833  & 2 & 271 & 6.76 \\
\textbf{Test} & 4055 & 3  & 327 &   12.4  \\
\bottomrule
\end{tabular}
\caption{Train and validation data summary}
\label{tab:data-stats}
\end{table}

\subsubsection{Conference and Journals}
\begin{figure}
    \centering
    \includegraphics[scale=0.35]{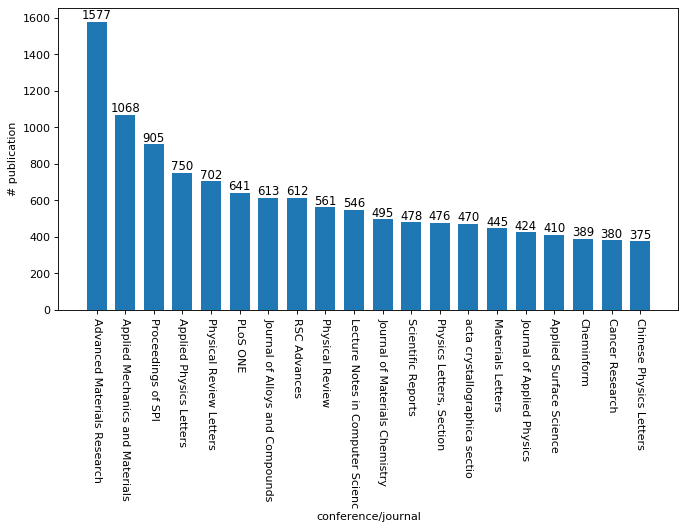}
    \caption{Paper frequency distribution by Conference/Journal }
    \label{fig:conference}
\end{figure}

Academic conferences are symposiums which researchers attend to present their findings and hear about the latest work in their field of interest. In Fig. \ref{fig:conference}, we have illustrated the frequency distribution of top 20 conferences/journals where authors have published their work. We plot the top 20 conferences/journals (by their publication count) on the x-axis and plot the publication count on y-axis. From this data, we can see vividly that conferences and journals in Applied Mechanics/Materials, Applied Physics and Bioinformatics are popular among the authors. This is validated by the keyword frequency distribution graph in Fig. \ref{fig:keyword} too where we see the top keywords pertaining to topics in these very fields. 

\subsubsection{Keywords}
\begin{figure}
    \centering
    \includegraphics[scale=0.35]{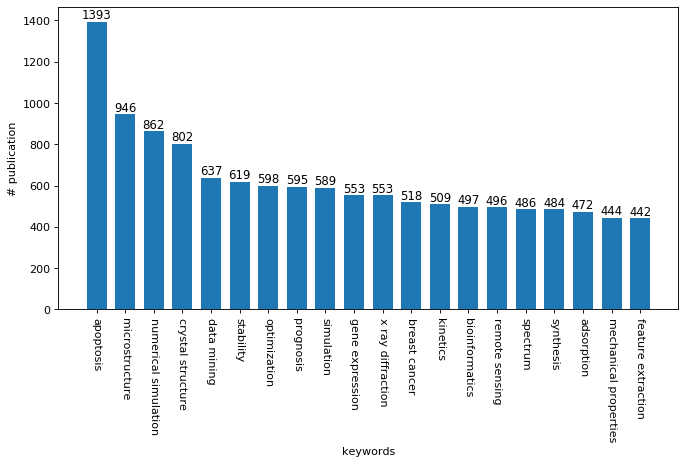}
    \caption{Keyword frequency distribution}
    \label{fig:keyword}
\end{figure}

  Effective keywords of an article portray an accurate representation of what an author wants to publish. Many-a-times, in the first glance, we look at the topic, keywords and abstract to get an idea about the research context of a publication. Fig. \ref{fig:keyword} illustrates the frequency distribution of top 20 keywords (by count) in the publications of the training dataset. We plot the top 20 selected keywords (by count) on the x-axis and have their counts plotted on the y-axis. This primarily gives us an idea about the different genres/topics of research where authors have published their work in. It can be seen that many of the publications contain keywords pertaining to the domain of Material research, Applied Mechanics and Bioinformatics.

\subsubsection{Year}
\begin{figure}
    \centering
    \includegraphics[scale=0.6]{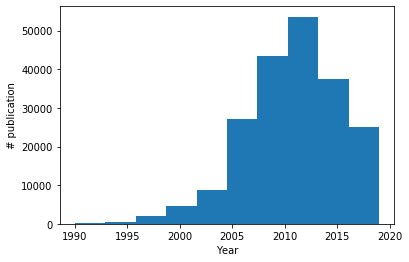}
    \caption{Publication frequency distribution by year}
    \label{fig:year}
\end{figure}

In Fig. \ref{fig:year}, we illustrate the publication frequency distribution by year. We see that our dataset consists of publications throughout the years from 1995 to 2019, with a majority of them being published between 2007-2017. This emphasizes on the recency of the dataset and better robustness to the present scenario. We plot years on the x-axis and the publication count of that year on the y-axis.

\subsubsection{Author name and affiliation}
\begin{figure}
    \centering
    \includegraphics[scale=0.6]{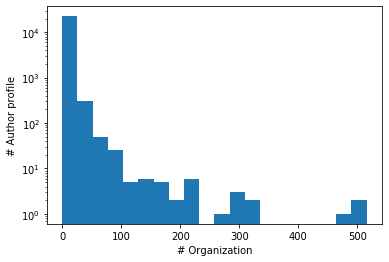}
    \caption{Author profile frequency distribution by organization count}
    \label{fig:auth_org}
\end{figure}

In Fig. \ref{fig:auth_org}, we illustrate the distribution of author profile count against the count of distinct organizations. More formally, we have recorded the number of author profiles on the y-axis who have been in corresponding number of distinct organizations on the x-axis. The graph shown in Fig. \ref{fig:auth_org} shows that there are many author profiles who have switched across organizations in their career which in turn strengthens the claim that many authors move across different organizations/places to cater to their research interests.

\section{Methods}
\subsection{Problem Formulation} \label{formuation}
\begin{figure*}
    \centering
    \includegraphics[scale=0.5]{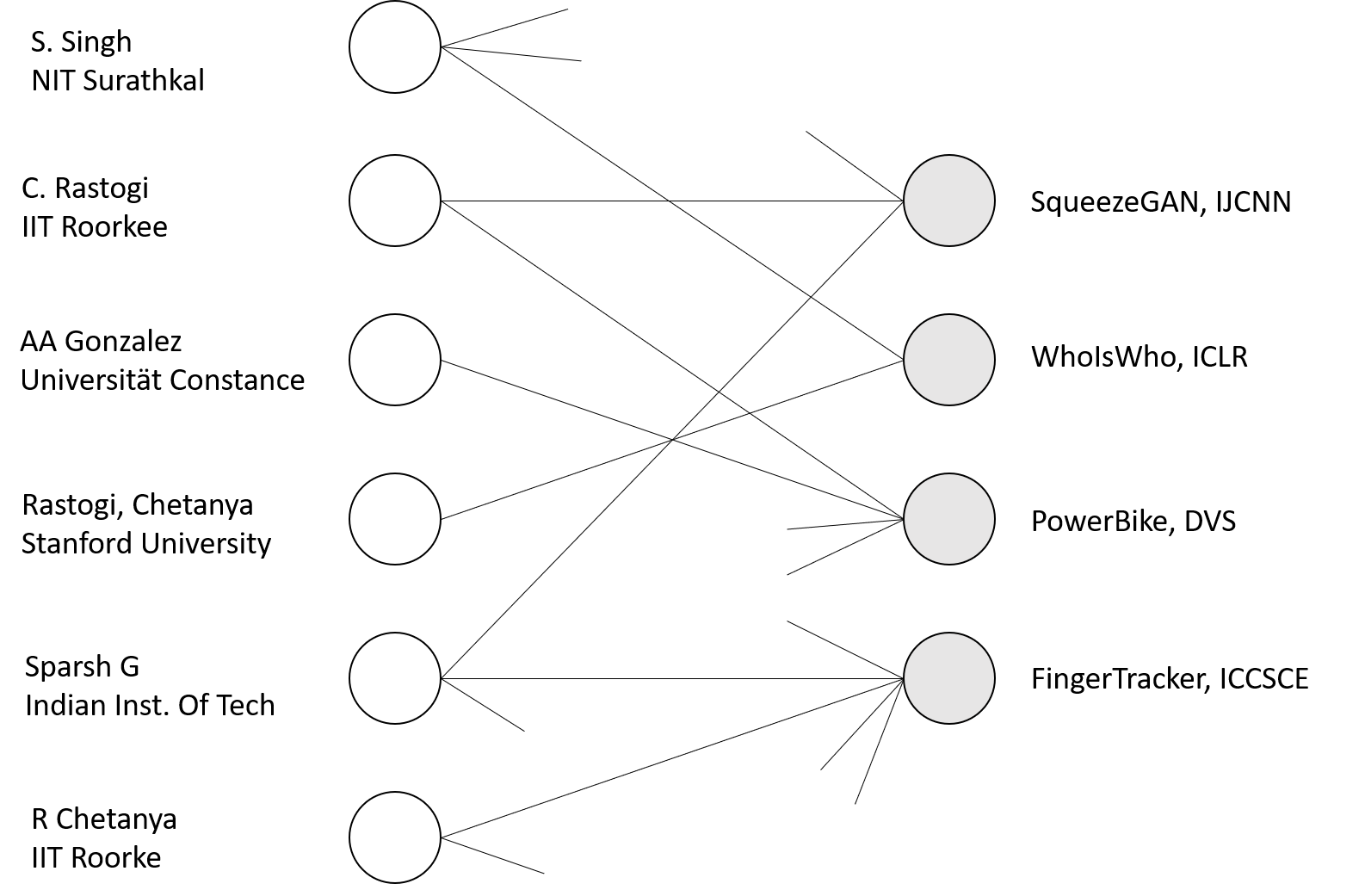}
    \caption{Bipartite graph of author node and publications}
    \label{fig:graph-problem}
\end{figure*}

We formulate the problem of author name disambiguation as finding similarity between nodes in a bipartite graph. Given a set of publications and their respective co-authors, we construct a bipartite graph as shown in Fig \ref{fig:graph-problem}.

Formally, given a set of publications $\mathcal{P}$, we construct a bipartite graph $\mathcal{G}$ as follows:
\begin{align*}
    \mathcal{G} &= (U, V, E) \\
    U &= \{ a | a \in p.authors \text{ and } p \in \mathcal{P}\} \\
    V &= \mathcal{P} \\
    E &= \{(a, p) | a \in p.authors \text{ and } p \in \mathcal{P}\} \\
\end{align*}

Now, we define the task of author name disambiguation as clustering the nodes with same author name in $U$ based on some node similarity function $AUTHOR\_SIM$. The clustering algorithm is shown in \ref{algo:clustering}.
\begin{algorithm}
\KwInput{$\mathcal{P}$: Set of publications (partitioned by author name) to cluster according to author profiles, $AUTHOR\_SIM$: author node similarity function}
\KwOutput{Cluster of publications according to author profile}
Clusters = \{\}

\ForEach{$p \in \mathcal{P}$}{
    Find cluster in Clusters with greatest similarity $s$ with author of interest in $p$ using the similarity function $AUTHOR\_SIM$\\
    If $s > \theta$, add $p$ to the maximum similarity cluster else create a new cluster
}
return Clusters

\caption{Clustering author-org nodes based on similarity function}
\label{algo:clustering}
\end{algorithm}
We analyse the behavior of various node similarity functions $AUTHOR\_SIM$ based on random walks (section \ref{sim_methods}), node embedding (section \ref{emb_methods}), and graph convolution networks (section \ref{convolution}).

\subsection{Evaluation} \label{eval}
The evaluation metric used in the task is macro-averaged F-1 score defined as below:
$$
\text{PairwisePrecision} = \frac{\#Pairs Correctly Predicted To Same Author}{\#Total Pairs Predicted To Same Author}
$$
$$
\text{PairwiseRecall} = \frac{\#\text{Pairs Correctly Predicted  To Same Author}}{\#\text{Total Pairs To Same Author}}
$$
$$
\text{Pairwise }F_1 = \frac{2 \times \text{PairwisePrecision} \times \text{PairwiseRecall}}{\text{PairwisePrecision} + \text{PairwiseRecall}}
$$
\subsection{Text based similarity (Baseline)}
We implement two baseline methods and study the performance of our system with respect to them. Both the baselines are described as follows with their performance summarised in Table \ref{table:Performance}:
\begin{enumerate}
    \item \textbf{ClusterByName}: 
    For the first baseline we combine all the authors with the same name under a single author profile.
    Formally, we define the $AUTHOR\_SIM$ function as follows:
    $$AUTHOR\_SIM(a1, a2) \gets a1.name = a2.name$$
    
    This provides a lower bound on any system's overall performance as it is likely that authors with the same name might be the same person (entity) and highly unlikely (though not impossible, since people use different forms of names) that authors with different names represent the same person (entity). As expected, the precision is quite low and the recall of the system is very high in this case.
    \item \textbf{ClusterByNameAndOrg}:  For the second baseline, we further fine-grained the author profiles with respect to their affiliation and name combined. Due to the highly noisy data, instead of doing a perfect match for organization name we make use of jaro-winkler similarity metric to match organizations.
    
    Formally, we define the $AUTHOR\_SIM$ function as follows:
    \begin{align*}
        AUTHOR\_SIM(a1, a2) &= a1.name == a2.name \\
        & \& \quad jaro(a1.org, a2.org) > 0.9
    \end{align*}
    
    Since it is highly unlikely that authors with the same name are affiliated with the same organization, this baseline ensures that we do not cluster different author profiles together but runs a risk of creating multiple profiles for a single author. The performance of the system degraded with this baseline which implies that authors frequently change affiliation over their lifetime which is validated from the author affiliation statistic in Fig. \ref{fig:auth_org}.
\end{enumerate}
\subsection{Random Walk based similarity}\label{sim_methods}
Random walk with restart (RWR) provides a good
relevance score between two nodes in a graph,
and it has been successfully used in numerous settings,
like automatic captioning of images, generalizations to
the "connection subgraphs", personalized PageRank, and
many more. Hence we use a slightly modified version of the RWR algorithm shown in algorithm \ref{algo:rwr_with_merge} to find and merge similar author nodes. In our version, the similar nodes are merged on the go after each random walk so that further iterations can benefit from the results of previous iterations.
\begin{algorithm}
\KwInput{$G$: Bipartite graph of authors and publications, $\alpha$: restart probability, $N$: max number of epochs, $W$: Random walk length, $T$: Threshold of visit count for merge}
\KwOutput{G with disambiguated nodes merged}
\For{$epoch \gets 1$ \KwTo $N$}{
    \ForEach{$authorNode \in G.V$}{
    visitCount $\gets$ \{\} \\
    $startNode$ $\gets$ $authorNode$
    $l \gets 0$ \\
    \While{$l < W$}{
        \If{$random < alpha$}{
        $authorNode \gets startNode$
        }
        \Else
        {
        sample a random neighbor $\mathbf{pubNode}$ of $authorNode$ \\
        sample a random neighbor $\mathbf{coAuthorNode}$ of $\mathbf{pubNode}$\\
        $authorNode \gets coAuthorNode$
        
        }
        $visitCount[authorNode] += 1$
    }
    Merge $startNode$ with nodes with $visitCount[node] > T$ if same is similar. 
}
}
\caption{Random walk based node merging}
\label{algo:rwr_with_merge}
\end{algorithm}
Formally, we define the $AUTHOR\_SIM$ function as follows:
    \begin{align*}
        AUTHOR\_SIM(a1, a2) &= a1.name == a2.name \\
        & \& \quad RWRVisitCount(a1, a2) > 0
    \end{align*}
Unlike the other similarity functions explored, in this method we update the graph online as we find similar nodes using the RWR visit count.
\subsection{Transductive embedding based similarity} \label{trans} \label{emb_methods}
Node Embeddings have been successful in many graph classification and clustering tasks and hence we explore both transductive and inductive embedding methods to define the author node similarity function. Inductive learning methods are described in next section under the graph convolution methods. In this section, we describe a popular transductive embedding method Node2Vec.

Node2Vec framework learns low-dimensional representations for nodes in a graph by optimizing a neighborhood preserving objective. The objective is flexible, and the algorithm accommodates for various definitions of network neighborhoods by simulating biased random walks. The two main user-defined hyperparameters $p$ and $q$ stand for the return and in-out hyperparameters respectively. The return parameter $p$ controls the probability of the walk staying inwards, revisiting the nodes again (exploitation); whereas the inout parameter $q$ controls the probability of the walk going farther out to explore other nodes (exploration).

In our approach, we run the Node2Vec algorithm on the bipartite graph $\mathcal{G}$ defined in section \ref{formuation}. In our setting, we run Node2Vec with the length of the walks set at 10, number of epochs set at 20 and p and q parameters set at 1.
\\
After running the Node2Vec algorithm, we derive the node embeddings $EMB$ of all the paper and author nodes of our graph.
Then we define the $AUTHOR\_SIM$ function as follows:
    \begin{align*}
        AUTHOR\_SIM(a1, a2) &= a1.name == a2.name \\
        & \& \quad cosine(EMB(a1), EMB(a2)) > \theta
    \end{align*}
    where $\theta$ is a user defined threshold.
\\
The intuition behind using Node2Vec embeddings to express the author and paper nodes is that Node2Vec leverages the inbuilt graphical properties of a graph by running multiple random walks across the nodes. Hence, running Node2Vec on the given Bipartite graph will result in placing similar author (author-organization) nodes together in the walks. The author (author-organization) nodes which occur together in multiple random walks will eventually get very similar embeddings by optimizing over the loss function of Node2Vec. Intuitively, this means that the author ((author-organization)) nodes having similar embeddings might belong to the same author, and hence, should be coalesced together; conditioned on some defined clustering threshold. We tabulate the results from this approach using different values for the threshold in Table \ref{table:n2vec}.

\subsection{Graph Neural Networks} \label{convolution}
Graph neural networks have been very successful in a variety of graph and node classification tasks as they can learn representation of the node incorporating both graph features and node features. Hence we use GNN to learn the $AUTHOR\_SIM$ function using both supervised and unsupervised setting. The initial node features used in these methods is described below:

\subsubsection{Features:} To take into account the meta-information of the publications, we make use of the various fields that accurately identify a publication. As mentioned in section \ref{sec:data}, we analyse all the fields and define the following features that are further used in all the graph neural network based approaches:
\begin{itemize}
    \item Title: Titles convey a very precise and specific information which is very unique w.r.t each publication. Therefore to incorporate the information contained in the title, we generate 100-dimensional Doc2Vec embeddings \cite{le2014distributed} which are obtained by training over the entire corpus of titles.
    \item Abstract: As in the case of a title, abstract also contain crucial information but unlike the title, they are at a higher level and in some sense convey the broader area to which the publication is related. Like titles, we generate 100-dimensional Doc2Vec embeddings \cite{le2014distributed} which are obtained by training over the entire corpus of abstracts.
    \item Year: We generate standardized year number for each paper with respect to the starting year number observed in the year distribution of the training corpus. We then use the standardized year number directly as a feature.
    \item Organization: Inspired by Name2Vec \cite{foxcroft2019name2vec}, we generate 100-dimensional embeddings of organization using Doc2Vec where each organization is represented as a document for character bigrams and trained over the whole corpus. 
\end{itemize}

To summarize, generate separate Doc2Vec embeddings \cite{le2014distributed} for \textit{abstract} and \textit{title} fields, each in a 100-dimensional space. Also to account for the activity of an author in the temporal space, we make use of the \textit{year} field and standardize it w.r.t a starting year. Similarly, we embed the \textit{org} field in a 100-dimensional space using Name2Vec \cite{name2vec}. Also we experimented with two different aggregation methods for combining feature information across nodes. First we projected all the individual features in a latent space and then combined them(sum) whereas in the second method we first combined all the features(concatenate) and then projected the combined feature space to a latent space. In our experiments we noticed that the latter approach (concatenate and project) performed better and hence we report all the results using this method. 

\subsubsection{Unsupervised Similarity Function}
In the unsupervised setting, given the author-publication bipartite graph $\mathcal{G}$, we want to learn embeddings for the nodes such that nodes close in the graph are more similar than those far away. The hypothesis here is that this will lead to node representations such that nodes belonging to same author profile will have similar embeddings when they are in close neighborhood as well as when they are in different components. Formally, given the initial node features $\mathbf{x}$, we calculate the embeddings $\mathbf{z}$ as follows:

\begin{align*}
    \mathbf{h}^0 &= \mathbf{x} \\
    \mathbf{h}^l &= \mathbf{GNN}(\mathbf{h}^l-1) \\ 
    \mathbf{z} &= \mathbf{h}^L \\
\end{align*}
We use the GNN described in PinSage \cite{ying2018graph} using neighborhood sampling to be able to apply this method on large academic graphs. To train the model to learn similar embeddings for nodes in close vicinity and dissimilar embeddings for faraway nodes, we used hinge loss as follows:
\begin{equation}
    \mathcal{L} = max(0, \mathbf{z}_{src}\mathbf{z}_{dst} -  \mathbf{z}_{src}\mathbf{z}_{dst\_neg} - \delta)
\end{equation}
Now, we define the $AUTHOR\_SIM$ function as follows:
    \begin{align*}
        AUTHOR\_SIM(a1, a2) &= a1.name == a2.name \\
        & \& \quad \mathbf{z}_{a1}\mathbf{z}_{a2} > \theta
    \end{align*}
    where $\theta$ is a user defined threshold.
The results of the model is shown in Table \ref{table:Performance}.
\subsubsection{GNN based Supervised Similarity Function}
\begin{figure*}
    \centering
    \includegraphics[scale=0.45]{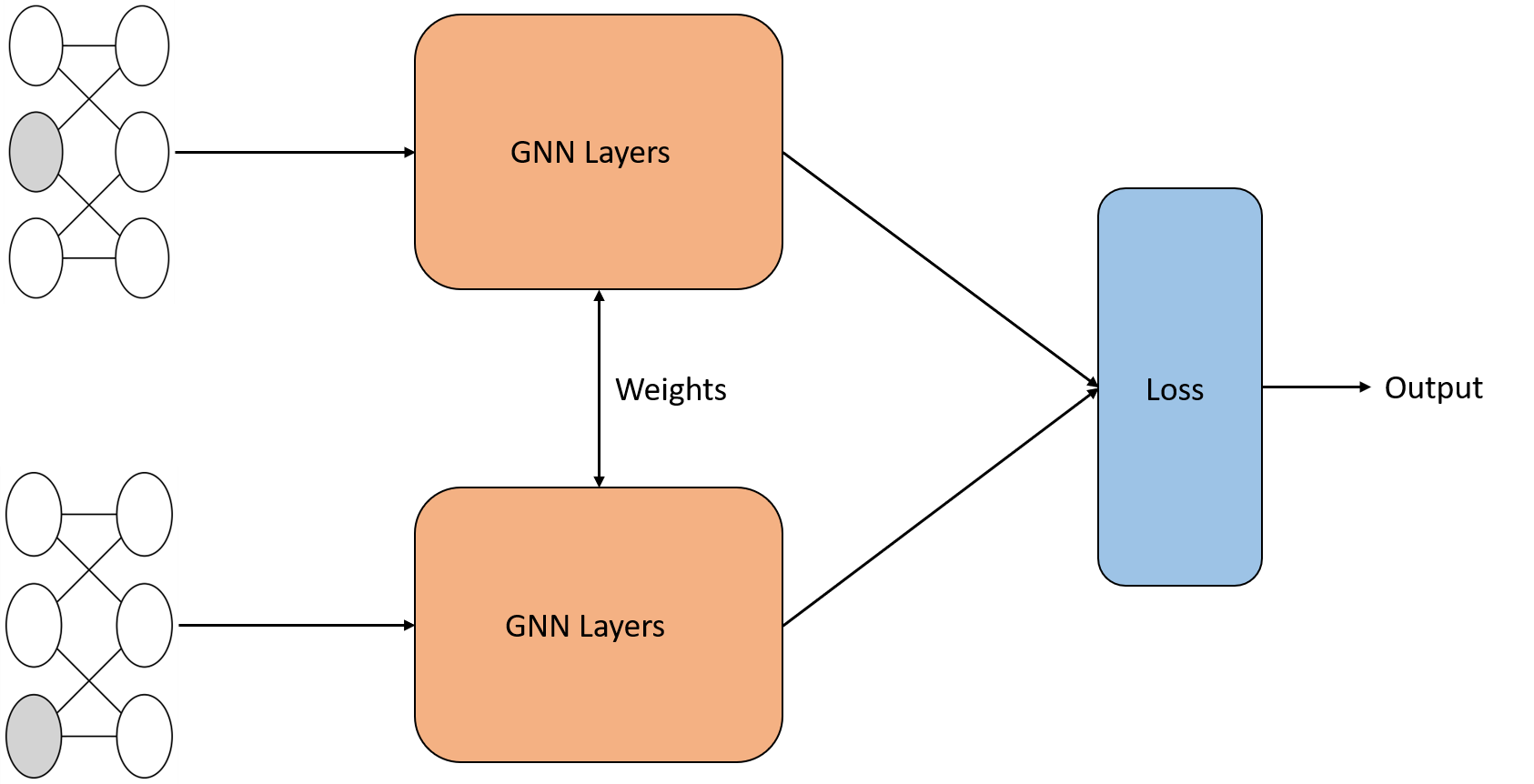}
    \caption{Architecture of network for supervised training}
    \label{fig:supervised-gnn}
\end{figure*}

In the supervised setting, we first create a dataset of pairs of author nodes consisting of pairs which are similar (belonging to the same author profile) and pairs which are dissimilar (belonging to different author profile with same or different name).
We then use a Siamese network $\mathcal{F}$ with negative log likelihood loss to learn the weights of the network (shown in \ref{fig:supervised-gnn}).
We then define the $AUTHOR\_SIM$ function as follows:
    \begin{align*}
        AUTHOR\_SIM(a1, a2) &= a1.name == a2.name \\
        & \& \quad \mathcal{F}(a1, a2) > \theta
    \end{align*}
    where $\theta$ is a user defined threshold.
We use different variations of the GNN architecture, results of which is shown in \ref{table:Performance}.
\subsubsection{MLP based Supervised Similarity Function}
To explore the effect of using graph features in finding similar nodes, we also train a model with only fully connected layers instead of the GNN layers in the above network. The results of the model is shown in \ref{table:Performance}.
\section{Results and Analysis} \label{results-analysis}
\begin{table}
\begin{tabular}{@{}lrrr@{}}
\toprule
\textbf{Method}              & \multicolumn{1}{l}{\textbf{pP}} & \multicolumn{1}{l}{\textbf{pR}} & \multicolumn{1}{l}{\textbf{pF}} \\ \midrule
\textbf{ClusterByName}       & 0.14                            & 1.00                            & 0.25                            \\
\textbf{ClusterByNameAndOrg} & 0.25                            & 0.12                            & 0.17                            \\ \midrule
\textbf{RWR-Merge} & 0.30  & 0.17 & 0.22 \\
\midrule
\textbf{Node2Vec} & 0.27 & 0.11 & 0.16 \\
\midrule
\textbf{Supervised-PinSage} & 0.22 & 0.26 & 0.24 \\
\textbf{Supervised-GraphSage} & 0.20 & 0.27 & 0.23\\
\textbf{Supervised-GCN} & 0.14 & 0.72 & 0.24  \\
\textbf{Supervised-MLP} & 0.14 & 0.28 & 0.18  \\
\midrule
\textbf{Unsupervised-Pinsage}  & 0.12 & 0.80 & 0.21\\
\bottomrule
\end{tabular}
\caption{Performance of baseline methods}
\label{table:Performance}
\end{table}
\subsection{RWR-Merge}
Table \ref{table:Sample_output} shows some of the sample nodes that were correctly identified and merged together by just using the network structure. Since we are only merging the nodes if both the nodes have the same name, this method was supposed to perform better on the pairwise precision metric without any guarantees on recall.
\subsubsection{Error Analysis:} We expected that this method to have a high pairwise precision but didn't get the desired results as shown in Table \ref{table:Performance}. On the careful inspection of the merged nodes, we found that the precision suffered because a high number of author nodes didn't had the \textit{org} info associated with them. This might have lead to a wrong initialization at the beginning of the algorithm as two different authors with same name but no \textit{org} were already provided as a single identity to the algorithm. Moreover, as the graph was highly disconnected, the algorithm had no chance of merging two nodes that were split across two or more connected components. This can be seen in the low pairwise recall values. 

\begin{table}
\begin{tabular}{@{}lp{0.82\linewidth}@{}}
\toprule
\textbf{Name}              & \multicolumn{1}{l}{\textbf{(Name,Org)}} \\
\midrule
\textbf{m\_giffels}       &  (m\_giffels, cern) \\ & (m\_giffels, rwth)    \\
\midrule 
\textbf{t\_dahms} & (t\_dahms,laboratoire leprinceringuet ecole polytechnique) \\ & (t\_dahms,cern european organization for nuclear research) \\
\midrule 
\textbf{e\_yildirim} & (e\_yildirim,enrico fermi institute) \\ & (e\_yildirim,desy) \\
\midrule 
\textbf{m\_k\_jha} & (m\_k\_jha,purdue university) \\ & (m\_k\_jha,university of puerto rico) \\
\bottomrule
\end{tabular}
\caption{Sample clusters produced by RWR-Merge}
\label{table:Sample_output}
\end{table}

\subsection{Node2Vec}
Table \ref{table:n2vec} tabulates precision, recall and F1 scores for different clustering thresholds used for clustering the Node2Vec vectors, as explained in Section \ref{trans}. The precision, recall and F1 scores are calculated according to the evaluate metrics defined in Section \ref{eval}. 
\subsubsection{Error Analysis} One of the samples classified correctly by the Node2Vec model (true positive) is for the author name 'Alessandro Giuliani' where two distinct author profiles are identified, the first containing papers with paper ids '5HWAan4P' ('A recursive network approach can identify constitutive regulatory circuits in gene expression data') and 'I7KqbI7a' ('Medical Data Analysis, Third International Symposium, ISMDA 2002, Rome, Italy, October 8-11, 2002, Proceedings') and the second containing papers with paper ids 'tznTWpXP' ('Multifractal characterization of protein contact networks') and 'fbcIaJuu' ('A generative model for protein contact networks'). Similarly, among the False positive examples, we have author name 'Yan Liang', for whom we cluster various papers of paperids 'zwzfkKwL' ('Track initiation algorithm based on Hough transform and clustering'), 'rZGsx5cX' ('Space-time linear dispersion codes based on optimal algorithms'), 'rl9MJQHl' ('Co-ordinative stock management system for permissible storage in VMI pattern') under the same author profile.
\begin{table}
\begin{tabular}{@{}lrrr@{}}
\toprule
\textbf{Similarity threshold}    & \multicolumn{1}{p{0.13\linewidth}}{\textbf{pP}} & \multicolumn{1}{p{0.14\linewidth}}{\textbf{pR}} & \multicolumn{1}{p{0.12\linewidth}}{\textbf{pF1}} \\ \midrule
\textbf{0.0} & 0.14 & 0.97 & 0.25    \\
\textbf{0.5} & 0.27 & 0.11 & 0.16    \\
\textbf{0.8} & 0.27 & 0.11 & 0.15   \\
\textbf{0.95} & 0.26 & 0.10 & 0.15   \\
\bottomrule
\end{tabular}
\caption{Node2Vec evaluation on different clustering thresholds}
\label{table:n2vec}
\end{table}

\subsection{Unsupervised GNN embeddings}
The results for unsupervised PinSage algorithm are shown in Table \ref{table:Performance}. This method had very high pairwise recall at the expense of pairwise precision. The method did extremely well in identifying similar author nodes that were scattered across multiple connected components of the bipartite graph but failed to capture fine-grained distinction between different author nodes with the same name.
\subsubsection{Error Analysis:} The unsupervised version of PinSage overcame the problem of RWR as it defined the similarity function which could assign non-trivial values to any two nodes in the entire graph. Due to this the algorithm successfully identified similar nodes across multiple connected components on the basis of the graph structure(like node degree, egonet, etc). However, the algorithm couldn't discriminate across distinct author profiles due to lack of supervision. This was evident from the fact that the final clustering obtained from this method had 2 clusters for each author name, one with high degree(high publications) and the other group with low degree. 

\subsection{Supervised GNN/MLP embeddings}
We used different GNN architectures to study the variation of performance depending upon the network. We also conducted an ablation study using only FC layers over the node features to study the effect of incorporating features of the neighbors. The results of different experiments in shown in table \ref{table:Performance}. We expect the MLP to perform poorly when compared to GNN layers and expect that the supervised setting will perform better than the unsupervised model above.

\subsubsection{Error Analysis}:
Since in the MLP architecture we are training only on author node features, i.e., the embedding of the organization, we expect the results to be similar to the baseline model where we clustered nodes based on the name and organization and this is indeed the case as can be seen from the table \ref{table:Performance}.

As compared to the unsupervised GNN embeddings, the supervised architecture is expected to perform better as the labels are directly fed into the system allowing the network that nodes in different components can also be similar and hence bias the network to look more into node features like the abstract and title of the publications. We observe that the different GNN architecture like GCN \cite{kipf2016semi}, GraphSage \cite{hamilton2017inductive}, PinSage \cite{ying2018graph} perform similarly on this task which is an interesting line to explore in the future.

Also, we have also observed that while the embedding similarities are quite skewed in the unsupervised setting rendering the model immune to threshold variation, the performance of the supervised model is dependent on the threshold giving a knob to tune recall and precision as required.

\section{Conclusion}
In this paper, we have thoroughly analysed the dataset hosted as a part of the Open Academic Graph WhoIsWho Track 1 and have implemented various techniques to specifically address the Author Name Disambiguation problem. Formally, we first represent the dataset in a Bipartite graph format containing two types of nodes: author and paper. We then define different flavours of the author similarity function to cluster the author nodes with same author name together. We experiment these different author similarity functions with (1) Text Based similarity (2) Random Walk based similarity (3) Transductive embedding based similarity and (4) Graph Neural Network methods and record results for the same. We conduct extensive quantitative and qualitative analysis of our dataset and graph, run several offline experiments with different combinations of Graph-based approach and author similarity functions and report the results. We observe that random walk based methods have high precision but low recall (as we cluster nodes conservatively) whereas embedding based methods in general have low precision and high recall (due to nodes across components being clustered together).

\section{Future Work}
We applied several architectures and learning paradigms to solve the problem of Author name Disambiguation and did a rigorous error analysis on these methods. Based on the results, one straightforward extension is to combine the RWR method along with other supervised learning techniques. This is because RWR can provide with a good starting point by aggregating some nodes which can result in high accuracy and low training time for these networks. Another area to focus is the tuning of the hyper-parameters to achieve a model of optimum performance as the models have shown significant promise in the initial experiments conducted by us.


\begin{acks}
We would like to thank Michele Catasta for guidance and consistent help throughout the course of the project and the generous google compute credits.
\end{acks}
\bibliographystyle{ACM-Reference-Format}
\bibliography{sample-base}


\begin{thebibliography}{29}


\ifx \showCODEN    \undefined \def \showCODEN     #1{\unskip}     \fi
\ifx \showDOI      \undefined \def \showDOI       #1{#1}\fi
\ifx \showISBNx    \undefined \def \showISBNx     #1{\unskip}     \fi
\ifx \showISBNxiii \undefined \def \showISBNxiii  #1{\unskip}     \fi
\ifx \showISSN     \undefined \def \showISSN      #1{\unskip}     \fi
\ifx \showLCCN     \undefined \def \showLCCN      #1{\unskip}     \fi
\ifx \shownote     \undefined \def \shownote      #1{#1}          \fi
\ifx \showarticletitle \undefined \def \showarticletitle #1{#1}   \fi
\ifx \showURL      \undefined \def \showURL       {\relax}        \fi
\providecommand\bibfield[2]{#2}
\providecommand\bibinfo[2]{#2}
\providecommand\natexlab[1]{#1}
\providecommand\showeprint[2][]{arXiv:#2}

\bibitem[\protect\citeauthoryear{Aggarwal and Subbian}{Aggarwal and
  Subbian}{2014}]%
        {aggarwal2014evolutionary}
\bibfield{author}{\bibinfo{person}{Charu Aggarwal} {and}
  \bibinfo{person}{Karthik Subbian}.} \bibinfo{year}{2014}\natexlab{}.
\newblock \showarticletitle{Evolutionary network analysis: A survey}.
\newblock \bibinfo{journal}{\emph{ACM Computing Surveys (CSUR)}}
  \bibinfo{volume}{47}, \bibinfo{number}{1} (\bibinfo{year}{2014}),
  \bibinfo{pages}{10}.
\newblock


\bibitem[\protect\citeauthoryear{Caron and van Eck}{Caron and van Eck}{2014}]%
        {caron2014large}
\bibfield{author}{\bibinfo{person}{Emiel Caron} {and} \bibinfo{person}{Nees~Jan
  van Eck}.} \bibinfo{year}{2014}\natexlab{}.
\newblock \showarticletitle{Large scale author name disambiguation using
  rule-based scoring and clustering}. In \bibinfo{booktitle}{\emph{Proceedings
  of the 19th international conference on science and technology indicators}}.
  CWTS-Leiden University Leiden, \bibinfo{pages}{79--86}.
\newblock


\bibitem[\protect\citeauthoryear{Chen}{Chen}{2019}]%
        {biendata}
\bibfield{author}{\bibinfo{person}{Bo Chen}.} \bibinfo{year}{2019}\natexlab{}.
\newblock \bibinfo{title}{{OAG-WhoIsWho Track 1}}.
\newblock
  \bibinfo{howpublished}{\url{https://www.biendata.com/competition/aminer2019/}}.
\newblock


\bibitem[\protect\citeauthoryear{De~Domenico, Omodei, and Arenas}{De~Domenico
  et~al\mbox{.}}{2016}]%
        {de2016quantifying}
\bibfield{author}{\bibinfo{person}{Manlio De~Domenico}, \bibinfo{person}{Elisa
  Omodei}, {and} \bibinfo{person}{Alex Arenas}.}
  \bibinfo{year}{2016}\natexlab{}.
\newblock \showarticletitle{Quantifying the diaspora of knowledge in the last
  century}.
\newblock \bibinfo{journal}{\emph{Applied network science}}
  \bibinfo{volume}{1}, \bibinfo{number}{1} (\bibinfo{year}{2016}),
  \bibinfo{pages}{15}.
\newblock


\bibitem[\protect\citeauthoryear{Foxcroft, d'Alessandro, and Antonie}{Foxcroft
  et~al\mbox{.}}{2019a}]%
        {name2vec}
\bibfield{author}{\bibinfo{person}{Jeremy Foxcroft}, \bibinfo{person}{Adrian
  d'Alessandro}, {and} \bibinfo{person}{Luiza Antonie}.}
  \bibinfo{year}{2019}\natexlab{a}.
\newblock \showarticletitle{Name2Vec: Personal Names Embeddings}. In
  \bibinfo{booktitle}{\emph{Advances in Artificial Intelligence}},
  \bibfield{editor}{\bibinfo{person}{Marie-Jean Meurs} {and}
  \bibinfo{person}{Frank Rudzicz}} (Eds.). \bibinfo{publisher}{Springer
  International Publishing}, \bibinfo{address}{Cham},
  \bibinfo{pages}{505--510}.
\newblock
\showISBNx{978-3-030-18305-9}


\bibitem[\protect\citeauthoryear{Foxcroft, d’Alessandro, and
  Antonie}{Foxcroft et~al\mbox{.}}{2019b}]%
        {foxcroft2019name2vec}
\bibfield{author}{\bibinfo{person}{Jeremy Foxcroft}, \bibinfo{person}{Adrian
  d’Alessandro}, {and} \bibinfo{person}{Luiza Antonie}.}
  \bibinfo{year}{2019}\natexlab{b}.
\newblock \showarticletitle{Name2Vec: Personal Names Embeddings}. In
  \bibinfo{booktitle}{\emph{Canadian Conference on Artificial Intelligence}}.
  Springer, \bibinfo{publisher}{Springer}, \bibinfo{pages}{505--510}.
\newblock


\bibitem[\protect\citeauthoryear{Hadiji, Mladenov, Bauckhage, and
  Kersting}{Hadiji et~al\mbox{.}}{2015}]%
        {hadiji2015computer}
\bibfield{author}{\bibinfo{person}{Fabian Hadiji}, \bibinfo{person}{Martin
  Mladenov}, \bibinfo{person}{Christian Bauckhage}, {and}
  \bibinfo{person}{Kristian Kersting}.} \bibinfo{year}{2015}\natexlab{}.
\newblock \showarticletitle{Computer science on the move: inferring migration
  regularities from the web via compressed label propagation}. In
  \bibinfo{booktitle}{\emph{Twenty-Fourth International Joint Conference on
  Artificial Intelligence}}.
\newblock


\bibitem[\protect\citeauthoryear{Hamilton, Ying, and Leskovec}{Hamilton
  et~al\mbox{.}}{2017}]%
        {hamilton2017inductive}
\bibfield{author}{\bibinfo{person}{Will Hamilton}, \bibinfo{person}{Zhitao
  Ying}, {and} \bibinfo{person}{Jure Leskovec}.}
  \bibinfo{year}{2017}\natexlab{}.
\newblock \showarticletitle{Inductive representation learning on large graphs}.
  In \bibinfo{booktitle}{\emph{Advances in Neural Information Processing
  Systems}}. \bibinfo{pages}{1024--1034}.
\newblock


\bibitem[\protect\citeauthoryear{Han, Giles, Zha, Li, and Tsioutsiouliklis}{Han
  et~al\mbox{.}}{2004}]%
        {Han:2004:TSL:996350.996419}
\bibfield{author}{\bibinfo{person}{Hui Han}, \bibinfo{person}{Lee Giles},
  \bibinfo{person}{Hongyuan Zha}, \bibinfo{person}{Cheng Li}, {and}
  \bibinfo{person}{Kostas Tsioutsiouliklis}.} \bibinfo{year}{2004}\natexlab{}.
\newblock \showarticletitle{Two Supervised Learning Approaches for Name
  Disambiguation in Author Citations}. In \bibinfo{booktitle}{\emph{Proceedings
  of the 4th ACM/IEEE-CS Joint Conference on Digital Libraries}}
  \emph{(\bibinfo{series}{JCDL '04})}. \bibinfo{publisher}{ACM},
  \bibinfo{address}{New York, NY, USA}, \bibinfo{pages}{296--305}.
\newblock
\showISBNx{1-58113-832-6}
\urldef\tempurl%
\url{https://doi.org/10.1145/996350.996419}
\showDOI{\tempurl}


\bibitem[\protect\citeauthoryear{Jindal, Singh, and Gadgil}{Jindal
  et~al\mbox{.}}{2023}]%
        {jindal2023classification}
\bibfield{author}{\bibinfo{person}{Akshat Jindal}, \bibinfo{person}{Shreya
  Singh}, {and} \bibinfo{person}{Soham Gadgil}.}
  \bibinfo{year}{2023}\natexlab{}.
\newblock \showarticletitle{Classification for everyone: Building geography
  agnostic models for fairer recognition}.
\newblock \bibinfo{journal}{\emph{arXiv preprint arXiv:2312.02957}}
  (\bibinfo{year}{2023}).
\newblock


\bibitem[\protect\citeauthoryear{Kipf and Welling}{Kipf and Welling}{2016}]%
        {kipf2016semi}
\bibfield{author}{\bibinfo{person}{Thomas~N Kipf} {and} \bibinfo{person}{Max
  Welling}.} \bibinfo{year}{2016}\natexlab{}.
\newblock \showarticletitle{Semi-supervised classification with graph
  convolutional networks}.
\newblock \bibinfo{journal}{\emph{arXiv preprint arXiv:1609.02907}}
  (\bibinfo{year}{2016}).
\newblock


\bibitem[\protect\citeauthoryear{Kumari and Singh}{Kumari and Singh}{2017}]%
        {kumari2017parallelization}
\bibfield{author}{\bibinfo{person}{SDGV~Akanksha Kumari} {and}
  \bibinfo{person}{Shreya Singh}.} \bibinfo{year}{2017}\natexlab{}.
\newblock \showarticletitle{Parallelization of alphabeta pruning algorithm for
  enhancing the two player games}.
\newblock \bibinfo{journal}{\emph{Int. J. Advances Electronics Comput. Sci}}
  \bibinfo{volume}{4} (\bibinfo{year}{2017}), \bibinfo{pages}{74--81}.
\newblock


\bibitem[\protect\citeauthoryear{Le and Mikolov}{Le and Mikolov}{2014}]%
        {le2014distributed}
\bibfield{author}{\bibinfo{person}{Quoc Le} {and} \bibinfo{person}{Tomas
  Mikolov}.} \bibinfo{year}{2014}\natexlab{}.
\newblock \showarticletitle{Distributed representations of sentences and
  documents}. In \bibinfo{booktitle}{\emph{International conference on machine
  learning}}. \bibinfo{pages}{1188--1196}.
\newblock


\bibitem[\protect\citeauthoryear{LLC}{LLC}{[n.d.]}]%
        {SCImago}
\bibfield{author}{\bibinfo{person}{MultiMedia LLC}.}
  \bibinfo{year}{[n.d.]}\natexlab{}.
\newblock \bibinfo{booktitle}{\emph{{MS Windows NT} Kernel Description}}.
\newblock
\urldef\tempurl%
\url{https://www.scimagojr.com/}
\showURL{%
\tempurl}


\bibitem[\protect\citeauthoryear{Ma, Wang, and Zhang}{Ma et~al\mbox{.}}{2019}]%
        {ma2019author}
\bibfield{author}{\bibinfo{person}{Xiao Ma}, \bibinfo{person}{Ranran Wang},
  {and} \bibinfo{person}{Yin Zhang}.} \bibinfo{year}{2019}\natexlab{}.
\newblock \showarticletitle{Author Name Disambiguation in Heterogeneous
  Academic Networks}. In \bibinfo{booktitle}{\emph{International Conference on
  Web Information Systems and Applications}}. Springer,
  \bibinfo{pages}{126--137}.
\newblock


\bibitem[\protect\citeauthoryear{Moed and Halevi}{Moed and Halevi}{2014}]%
        {moed2014bibliometric}
\bibfield{author}{\bibinfo{person}{Henk~F Moed} {and} \bibinfo{person}{Gali
  Halevi}.} \bibinfo{year}{2014}\natexlab{}.
\newblock \showarticletitle{A bibliometric approach to tracking international
  scientific migration}.
\newblock \bibinfo{journal}{\emph{Scientometrics}} \bibinfo{volume}{101},
  \bibinfo{number}{3} (\bibinfo{year}{2014}), \bibinfo{pages}{1987--2001}.
\newblock


\bibitem[\protect\citeauthoryear{Mohammed~Abdulla, Singh, and
  Borar}{Mohammed~Abdulla et~al\mbox{.}}{2019}]%
        {10.1145/3308560.3316599}
\bibfield{author}{\bibinfo{person}{G Mohammed~Abdulla}, \bibinfo{person}{Shreya
  Singh}, {and} \bibinfo{person}{Sumit Borar}.}
  \bibinfo{year}{2019}\natexlab{}.
\newblock \showarticletitle{Shop Your Right Size: A System for Recommending
  Sizes for Fashion Products}. In \bibinfo{booktitle}{\emph{Companion
  Proceedings of The 2019 World Wide Web Conference}}
  \emph{(\bibinfo{series}{WWW '19})}. \bibinfo{publisher}{Association for
  Computing Machinery}, \bibinfo{address}{New York, NY, USA},
  \bibinfo{pages}{327–334}.
\newblock
\showISBNx{9781450366755}
\urldef\tempurl%
\url{https://doi.org/10.1145/3308560.3316599}
\showDOI{\tempurl}


\bibitem[\protect\citeauthoryear{Orman, Labatut, and Naskali}{Orman
  et~al\mbox{.}}{2017}]%
        {orman2017exploring}
\bibfield{author}{\bibinfo{person}{G{\"u}nce~Keziban Orman},
  \bibinfo{person}{Vincent Labatut}, {and} \bibinfo{person}{Ahmet~Teoman
  Naskali}.} \bibinfo{year}{2017}\natexlab{}.
\newblock \showarticletitle{Exploring the evolution of node neighborhoods in
  Dynamic Networks}.
\newblock \bibinfo{journal}{\emph{Physica A: Statistical Mechanics and its
  Applications}}  \bibinfo{volume}{482} (\bibinfo{year}{2017}),
  \bibinfo{pages}{375--391}.
\newblock


\bibitem[\protect\citeauthoryear{Robinson-Garcia, Sugimoto, Murray,
  Yegros-Yegros, Larivi{\`e}re, and Costas}{Robinson-Garcia
  et~al\mbox{.}}{2019}]%
        {robinson2019many}
\bibfield{author}{\bibinfo{person}{Nicol{\'a}s Robinson-Garcia},
  \bibinfo{person}{Cassidy~R Sugimoto}, \bibinfo{person}{Dakota Murray},
  \bibinfo{person}{Alfredo Yegros-Yegros}, \bibinfo{person}{Vincent
  Larivi{\`e}re}, {and} \bibinfo{person}{Rodrigo Costas}.}
  \bibinfo{year}{2019}\natexlab{}.
\newblock \showarticletitle{The many faces of mobility: Using bibliometric data
  to measure the movement of scientists}.
\newblock \bibinfo{journal}{\emph{Journal of Informetrics}}
  \bibinfo{volume}{13}, \bibinfo{number}{1} (\bibinfo{year}{2019}),
  \bibinfo{pages}{50--63}.
\newblock


\bibitem[\protect\citeauthoryear{Schulz, Mazloumian, Petersen, Penner, and
  Helbing}{Schulz et~al\mbox{.}}{2014}]%
        {schulz2014exploiting}
\bibfield{author}{\bibinfo{person}{Christian Schulz}, \bibinfo{person}{Amin
  Mazloumian}, \bibinfo{person}{Alexander~M Petersen}, \bibinfo{person}{Orion
  Penner}, {and} \bibinfo{person}{Dirk Helbing}.}
  \bibinfo{year}{2014}\natexlab{}.
\newblock \showarticletitle{Exploiting citation networks for large-scale author
  name disambiguation}.
\newblock \bibinfo{journal}{\emph{EPJ Data Science}} \bibinfo{volume}{3},
  \bibinfo{number}{1} (\bibinfo{year}{2014}), \bibinfo{pages}{11}.
\newblock


\bibitem[\protect\citeauthoryear{Sebastian, Singh, Manikanta, Ashwin, and
  Reddy}{Sebastian et~al\mbox{.}}{2019}]%
        {10.1007/978-981-10-8639-7_25}
\bibfield{author}{\bibinfo{person}{Abraham~Gerard Sebastian},
  \bibinfo{person}{Shreya Singh}, \bibinfo{person}{P.~B.~T. Manikanta},
  \bibinfo{person}{T.~S. Ashwin}, {and} \bibinfo{person}{G.~Ram~Mohana Reddy}.}
  \bibinfo{year}{2019}\natexlab{}.
\newblock \showarticletitle{Multimodal Group Activity State Detection for
  Classroom Response System Using Convolutional Neural Networks}. In
  \bibinfo{booktitle}{\emph{Recent Findings in Intelligent Computing
  Techniques}}, \bibfield{editor}{\bibinfo{person}{Pankaj~Kumar Sa},
  \bibinfo{person}{Sambit Bakshi}, \bibinfo{person}{Ioannis~K.
  Hatzilygeroudis}, {and} \bibinfo{person}{Manmath~Narayan Sahoo}} (Eds.).
  \bibinfo{publisher}{Springer Singapore}, \bibinfo{address}{Singapore},
  \bibinfo{pages}{245--251}.
\newblock
\showISBNx{978-981-10-8639-7}


\bibitem[\protect\citeauthoryear{Singh, Singh, Arora, and Borar}{Singh
  et~al\mbox{.}}{2019}]%
        {singh2019one}
\bibfield{author}{\bibinfo{person}{Loveperteek Singh}, \bibinfo{person}{Shreya
  Singh}, \bibinfo{person}{Sagar Arora}, {and} \bibinfo{person}{Sumit Borar}.}
  \bibinfo{year}{2019}\natexlab{}.
\newblock \showarticletitle{One embedding to do them all}.
\newblock \bibinfo{journal}{\emph{arXiv preprint arXiv:1906.12120}}
  (\bibinfo{year}{2019}).
\newblock


\bibitem[\protect\citeauthoryear{Singh, Abdulla, Borar, and Arora}{Singh
  et~al\mbox{.}}{2018}]%
        {singh2018footwear}
\bibfield{author}{\bibinfo{person}{Shreya Singh}, \bibinfo{person}{G~Mohammed
  Abdulla}, \bibinfo{person}{Sumit Borar}, {and} \bibinfo{person}{Sagar
  Arora}.} \bibinfo{year}{2018}\natexlab{}.
\newblock \showarticletitle{Footwear Size Recommendation System}.
\newblock \bibinfo{journal}{\emph{arXiv preprint arXiv:1806.11423}}
  (\bibinfo{year}{2018}).
\newblock


\bibitem[\protect\citeauthoryear{Sinha, Shen, Song, Ma, Eide, Hsu, and
  Wang}{Sinha et~al\mbox{.}}{2015}]%
        {sinha2015overview}
\bibfield{author}{\bibinfo{person}{Arnab Sinha}, \bibinfo{person}{Zhihong
  Shen}, \bibinfo{person}{Yang Song}, \bibinfo{person}{Hao Ma},
  \bibinfo{person}{Darrin Eide}, \bibinfo{person}{Bo-june~Paul Hsu}, {and}
  \bibinfo{person}{Kuansan Wang}.} \bibinfo{year}{2015}\natexlab{}.
\newblock \showarticletitle{An overview of microsoft academic service (mas) and
  applications}. In \bibinfo{booktitle}{\emph{Proceedings of the 24th
  international conference on world wide web}}. ACM, \bibinfo{pages}{243--246}.
\newblock


\bibitem[\protect\citeauthoryear{Tang, Fong, Wang, and Zhang}{Tang
  et~al\mbox{.}}{2011}]%
        {tang2011unified}
\bibfield{author}{\bibinfo{person}{Jie Tang}, \bibinfo{person}{Alvis~CM Fong},
  \bibinfo{person}{Bo Wang}, {and} \bibinfo{person}{Jing Zhang}.}
  \bibinfo{year}{2011}\natexlab{}.
\newblock \showarticletitle{A unified probabilistic framework for name
  disambiguation in digital library}.
\newblock \bibinfo{journal}{\emph{IEEE Transactions on Knowledge and Data
  Engineering}} \bibinfo{volume}{24}, \bibinfo{number}{6}
  (\bibinfo{year}{2011}), \bibinfo{pages}{975--987}.
\newblock


\bibitem[\protect\citeauthoryear{Treeratpituk and Giles}{Treeratpituk and
  Giles}{2009}]%
        {treeratpituk2009disambiguating}
\bibfield{author}{\bibinfo{person}{Pucktada Treeratpituk} {and}
  \bibinfo{person}{C~Lee Giles}.} \bibinfo{year}{2009}\natexlab{}.
\newblock \showarticletitle{Disambiguating authors in academic publications
  using random forests}. In \bibinfo{booktitle}{\emph{Proceedings of the 9th
  ACM/IEEE-CS joint conference on Digital libraries}}. ACM,
  \bibinfo{pages}{39--48}.
\newblock


\bibitem[\protect\citeauthoryear{Ying, He, Chen, Eksombatchai, Hamilton, and
  Leskovec}{Ying et~al\mbox{.}}{2018}]%
        {ying2018graph}
\bibfield{author}{\bibinfo{person}{Rex Ying}, \bibinfo{person}{Ruining He},
  \bibinfo{person}{Kaifeng Chen}, \bibinfo{person}{Pong Eksombatchai},
  \bibinfo{person}{William~L Hamilton}, {and} \bibinfo{person}{Jure Leskovec}.}
  \bibinfo{year}{2018}\natexlab{}.
\newblock \showarticletitle{Graph convolutional neural networks for web-scale
  recommender systems}. In \bibinfo{booktitle}{\emph{Proceedings of the 24th
  ACM SIGKDD International Conference on Knowledge Discovery \& Data Mining}}.
  ACM, \bibinfo{pages}{974--983}.
\newblock


\bibitem[\protect\citeauthoryear{Zhang, Saha, and Al~Hasan}{Zhang
  et~al\mbox{.}}{2014}]%
        {zhang2014name}
\bibfield{author}{\bibinfo{person}{Baichuan Zhang},
  \bibinfo{person}{Tanay~Kumar Saha}, {and} \bibinfo{person}{Mohammad
  Al~Hasan}.} \bibinfo{year}{2014}\natexlab{}.
\newblock \showarticletitle{Name disambiguation from link data in a
  collaboration graph}. In \bibinfo{booktitle}{\emph{2014 IEEE/ACM
  International Conference on Advances in Social Networks Analysis and Mining
  (ASONAM 2014)}}. IEEE, \bibinfo{pages}{81--84}.
\newblock


\bibitem[\protect\citeauthoryear{Zhang, Zhang, Yao, and Tang}{Zhang
  et~al\mbox{.}}{2018}]%
        {Zhang:2018:NDA:3219819.3219859}
\bibfield{author}{\bibinfo{person}{Yutao Zhang}, \bibinfo{person}{Fanjin
  Zhang}, \bibinfo{person}{Peiran Yao}, {and} \bibinfo{person}{Jie Tang}.}
  \bibinfo{year}{2018}\natexlab{}.
\newblock \showarticletitle{Name Disambiguation in AMiner: Clustering,
  Maintenance, and Human in the Loop.}. In
  \bibinfo{booktitle}{\emph{Proceedings of the 24th ACM SIGKDD International
  Conference on Knowledge Discovery \&\#38; Data Mining}}
  \emph{(\bibinfo{series}{KDD '18})}. \bibinfo{publisher}{ACM},
  \bibinfo{address}{New York, NY, USA}, \bibinfo{pages}{1002--1011}.
\newblock
\showISBNx{978-1-4503-5552-0}
\urldef\tempurl%
\url{https://doi.org/10.1145/3219819.3219859}
\showDOI{\tempurl}


\end{thebibliography}

\appendix

\end{document}